\documentclass[11pt,oneside,letterpaper]{article}
\usepackage{geometry}                		%
\usepackage{graphicx}				%
\usepackage{amssymb}
\usepackage{physics}
\usepackage[colorlinks=false, urlcolor=blue, linkcolor=red]{hyperref}
\usepackage{cite}
\usepackage[symbol]{footmisc}

\title{\bf Direct, analytic solution for the electromagnetic vector potential in any gauge}
\author
{Kuo-Ho Yang \\
Department of Engineering and Physics, St.~Ambrose University,\\ 
Davenport, IA 52803 \\
E-mail: yangkuoho@sau.edu\\
Robert D. Nevels\\
Department of Electrical and Computer Engineering, Texas A\&M University,\\
College Station, TX 77843
}
\date{}					%
\begin{document}
\maketitle
\begin{abstract}
We derive an analytic solution for the electromagnetic vector potential in any gauge directly from Maxwell's equations for potentials for an arbitrary time-dependent charge-current distribution.  No gauge condition is used in the derivation.  Our solution for the vector potential has a gauge-invariant part and a gauge-dependent part.  The gauge-dependent part is related to the scalar potential. 
\end{abstract}
\section{Introduction}

To understand gauge transformations or gauge invariance in classical electrodynamics, it is essential that we have detailed knowledge of the vector and the scalar potentials.  In general, the potentials are solved from Maxwell's equations for potentials.  But, Maxwell's equations for potentials contain a gauge ambiguity.  If two sets of potentials are related by a gauge transformation, then both sets are solutions of the equations ({\it e.g.}~Refs.~\cite{Jackson-Okun-2001, Jackson-1999}).

Thus, to solve for a {\em unique} set of potentials, we add a constraint involving the vector or the scalar potential or both.  Such a constraint is called a gauge condition. Then, we solve for the potentials from the resulting equations after the gauge condition is applied.  But, each gauge condition generates its own equations for the potentials and hence its own mathematical challenges.  For example, both the scalar and the vector potentials in the Lorenz gauge can be solved simply.  But, the situation is not as straightforward for the Coulomb gauge.  Although the Coulomb-gauge scalar potential is easy to solve, it was only recently that we finally succeeded in obtaining analytic solutions for the vector potential in the Coulomb gauge for time-{\em dependent} charge and current densities \cite{Yang-gauge-1976, Brown-Crothers-1989, Rynne-Smith-Nevels-1991, Jackson-2002, Hnizdo-2004, Yang-gauge-2005, Heras-2007, Hnizdo-2007, Wun-Jen-2012, Yang-Nevels-2014, Yang-McDonald-2015, GTM-v-gauge-2022, Hnizdo-Vaman-2023}. 

In this paper, we derive an analytic solution for the vector potential ${\bf A}$ in any gauge for an arbitrary time-dependent  charge-current distribution.  The vector potential is derived {\em directly} from Maxwell's equations for potentials {\em without} the use of a gauge condition. Hence, the solution for ${\bf A}$ is universally valid for any gauge.  The vector potential has a gauge-invariant part and a gauge-dependent part.  The gauge-dependent part is related to the scalar potential $\Phi$.  We show that the fields generated by our potentials in any gauge are gauge invariant and {\em always} propagate with speed $c$ from physical charge and current densities.  In Appendix A, we solve the vector potential using Fourier transforms, and, in Appendix B, we solve the velocity-gauge vector potential directly from Maxwell's equations for potentials in the velocity gauge.

\section{Direct, analytic solution for the vector potential in any gauge from Maxwell's equations for potentials}

We consider localized charge and current densities, $\rho ({\bf r}, t)$ and ${\bf J} ({\bf r}, t)$, which are turned on at $t_0$.  The electric field ${\bf E}$, the magnetic field ${\bf B}$, and Maxwell's equations for potentials ${\bf A}$ and $\Phi$ are (in Gaussian units):
\begin{equation}
{\bf E}({\bf r}, t) = - \grad \Phi({\bf r}, t) - {1 \over c} {\partial {\bf A}({\bf r}, t) \over \partial t},
\qquad
{\bf B}({\bf r}, t) = \grad \times {\bf A}({\bf r}, t),
\label{eq-A1}
\end{equation}
\begin{equation}
\grad^2 \Phi ({\bf r}, t) + {1 \over c} {\partial \over \partial t} [ \grad \cdot {\bf A}({\bf r}, t) ] 
= - 4 \pi \rho ({\bf r}, t),
\label{eq-A2}
\end{equation}
\begin{equation}
\left( \grad^2 - {1 \over c^2} {\partial^2 \over \partial t^2}  \right) {\bf A}({\bf r}, t)  
= - { 4 \pi \over c } {\bf J}({\bf r}, t) 
  + \grad \left( \grad \cdot {\bf A}({\bf r}, t) + {1 \over c} {\partial \Phi ({\bf r}, t) \over \partial t} \right).
\label{eq-A3}
\end{equation}
We assume that there are no boundary surfaces anywhere.  We state that all quantities and operations in this paper are defined in the usual 3-dimensional space, {\it e.g.}, ${\bf r} = (x, y, z)$, ${\bf E} = (E_x, E_y, E_z)$, ${\grad} = (\partial/\partial x, \partial/\partial y, \partial/\partial z)$, {\it etc}. For a general n-dimensional solution to the wave equation, see Ref.~\cite{Barton-1989}.

To solve the vector potential ${\bf A}$ from eq.~(\ref{eq-A3}), we write ${\bf A} = {\bf A}_{1} + {\bf A}_{2}$ where ${\bf A}_{1}$ and ${\bf A}_{2}$ are solutions of their respective equations:
\begin{equation}
\left( \grad^2 - {1 \over c^2} {\partial^2 \over \partial t^2}  \right) {\bf A}_{1} 
= - {4 \pi \over c} {\bf J},
\label{eq-A4}
\end{equation}
\begin{equation}
\left( \grad^2 - {1 \over c^2} {\partial^2 \over \partial t^2}  \right) {\bf A}_{2} 
=  \grad \left[ \grad \cdot ({\bf A}_1 + {\bf A}_2) + {1 \over c} {\partial \Phi \over \partial t} \right].
\label{eq-A5}
\end{equation}
From eq.~(\ref{eq-A4}), it is clear that the solution for ${\bf A}_{1}$ is the $c$-retarded vector potential produced by the current density ${\bf J}$:
\begin{equation}
{\bf A}_{1} ({\bf r}, t) = {\bf A}_{c} ({\bf r}, t) 
= {1 \over c} \int {{\bf J}({\bf r'}, t - R/c) \over R} d^3r',
\qquad \qquad
R = |{\bf r} - {\bf r}'|.
\label{eq-A6}
\end{equation}

To solve for ${\bf A}_{2}$, we differentiate both sides of eq.~(\ref{eq-A5}) with respect to $t$ and use eq.~(\ref{eq-A2}) to eliminate $(\partial/\partial t)[\grad \cdot ({\bf A}_1 + {\bf A}_2)]$ to get,
\begin{equation}
\left( \grad^2 - {1 \over c^2} {\partial^2 \over \partial t^2}  \right)  \left( {\partial {\bf A}_{2} \over \partial t} \right)
= - 4 \pi c \grad \rho -  \left( \grad^2 - {1 \over c^2} {\partial^2 \over \partial t^2} \right) (c \grad \Phi).
\label{eq-A7}
\end{equation}
This equation can be solved in terms of the $c$-retarded scalar potential produced by the charge density $\rho$:
\begin{equation}
{\bf A}_{2} ({\bf r}, t) = c \grad \int \left[ \Phi_{c} ({\bf r}, t) - \Phi ({\bf r}, t) \right] dt, %
\label{eq-A8}
\end{equation}
\begin{equation}
\Phi_{c} ({\bf r}, t) = \int {\rho ({\bf r}', t - R/c) \over R}d^3r'.
\label{eq-A9}
\end{equation}
Thus, the full expression for the vector potential is:
\begin{equation}
{\bf A}  ({\bf r}, t) = {\bf A}_{c}  ({\bf r}, t) + c \grad \int \left[ \Phi_{c} ({\bf r}, t) - \Phi ({\bf r}, t) \right] dt.%
\label{eq-A10}
\end{equation}
Because eq.~({\ref{eq-A10}) is derived without the use of a gauge condition, it is valid for any gauge.  

We note that ${\bf A}_2$ in eq.~(\ref{eq-A8}) and consequently the full potential ${\bf A}$ in eq.~(\ref{eq-A10}) are only determined to within an additive time-independent gradient function of the form $\grad \Omega ({\bf r})$, which is set to be zero in this paper for brevity of discussions.  A similar situation also exists for the scalar potentials \cite{Hnizdo-2024}.

We also note that the vector potential in eq.~(\ref{eq-A10}) clearly satisfies eq.~(\ref{eq-A2}) for {\em any} $\Phi$:
\begin{eqnarray}
\grad^2 \Phi + {1 \over c} {\partial \over \partial t} ( \grad \cdot {\bf A} )
= \grad^2 \Phi + \left[ {1 \over c} {\partial \over \partial t} ( \grad \cdot {\bf A}_{c}) 
+ ( \grad^2 \Phi_{c} - \grad^2 \Phi)  \right]
\nonumber
\\
= \grad^2 \Phi_{c} -  {1 \over c^2} {\partial^2 \over \partial t^2} \Phi_{c}
= - 4 \pi \rho,
\qquad \qquad \qquad \ \,
\label{eq-A11}
\end{eqnarray}  
where we have used $\grad \cdot {\bf A}_{c} + c^{-1} \partial \Phi_{c} / \partial t = 0$.  

In Appendix A, we solve the vector potential using the method of Fourier transforms to make the mathematics more transparent.

\section{Potentials in the velocity gauge}

To apply eq.~(\ref{eq-A10}) to obtain the vector potential in a particular gauge, we first must find the scalar potential in that gauge.  For this purpose, we first start with a gauge condition and derive an equation for the scalar potential.  We then solve for the scalar potential and use it in eq.~(\ref{eq-A10}) to derive the vector potential.  Here, we use the velocity gauge as an example to show how this procedure works.

The velocity gauge with the parameter $v$ has the gauge condition \cite{Yang-gauge-1976, Brown-Crothers-1989, Jackson-2002, Yang-gauge-2005, Yang-McDonald-2015, GTM-v-gauge-2022}:
\begin{equation}
\grad \cdot {\bf A}^{(v)} + {c \over v^2} {\partial \Phi^{(v)} \over \partial t} = 0,
\label{eq-B1}
\end{equation}
where we set $v > 0$ by only considering the retarded solutions.    We use this gauge condition in eq.~(\ref{eq-A2}) to derive the equation for the scalar potential:
\begin{equation}
\left( \grad^2 - {1 \over v^2} {\partial^2 \over \partial t^2} \right) \Phi^{(v)} = - 4 \pi \rho.
\label{eq-B2}
\end{equation}
If we only consider the inhomogeneous solution, then the scalar potential propagates with speed $v$ from the charge density:
\begin{equation}
\Phi^{(v)}({\bf r}, t) = \int {\rho({\bf r}', t - R/v) \over R} d^3r', \qquad \qquad R = |{\bf r} - {\bf r}'|.
\label{eq-B3}
\end{equation}
According to eq.~(\ref{eq-A10}) the vector potential corresponding to this scalar potential is:
\begin{equation}
{\bf A}^{(v)} ({\bf r}, t) = {\bf A}_{c}({\bf r}, t)  + c \grad \int \left[ \Phi_{c} ({\bf r}, t) - \Phi^{(v)} ({\bf r}, t) \right] dt.
\label{eq-B4}
\end{equation}
In Appendix B, we solve the vector potential directly from Maxwell's equations for potentials in the velocity gauge and obtain the same answer.

This expression for the $v$-gauge vector potential was first derived by Yang \cite{Yang-gauge-1976, Yang-gauge-2005} using the arguments that electromagnetic fields are gauge invariant and propagate with speed $c$ from physical charge and current densities.  See also Ref.~\cite{Brown-Crothers-1989}, Appendix 3 for a derivation of eq.~(\ref{eq-B4}).

Let us construct a new scalar potential $\Phi^{({\rm new})}$ by combining the $v$-gauge potential $\Phi^{(v)}$ with an arbitrary scalar function $\xi ({\bf r}, t)$ as follows:
\begin{equation}
\Phi^{({\rm new})} = \Phi^{(v)} - \xi = \Phi^{(v)} - {1 \over c} {\partial \over \partial t} {\left( c \int \xi ({\bf r}, t) dt \right)}.
\label{eq-B5}
\end{equation}
Then, we use eq.~(\ref{eq-B5}) in (\ref{eq-A10}) to derive the corresponding new vector potential:
\begin{equation}
 {\bf A}^{({\rm new})} = {\bf A}_{c}  + c \grad \int \left[ \Phi_{c} - \left( \Phi^{(v)} - \xi \right) \right] dt
=  {\bf A}^{(v)} + \grad {\left( c \int \xi ({\bf r}, t) dt \right)}.
\label{eq-B6}
\end{equation}
Both sets of potentials, $(\Phi^{(v)}, {\bf A}^{(v)})$ and $(\Phi^{({\rm new})}, {\bf A}^{({\rm new})})$, are valid potentials because they both are solutions of Maxwell's equations for potentials eqs.~(\ref{eq-A2})-(\ref{eq-A3}). As a consequence, they generate the same electric and magnetic fields.  We note that $(\Phi^{({\rm new})}, {\bf A}^{({\rm new})})$ are related to $(\Phi^{(v)}, {\bf A}^{(v)})$ by a gauge transformation.

The potentials in the Lorenz gauge satisfying the boundary condition that they vanish at $|{\bf r}| \rightarrow \infty$ are obtained by setting $v = c$ in eqs.~(\ref{eq-B1})-\ref{eq-B4}): 
\begin{equation}
\Phi^{(L)} ({\bf r}, t) = \Phi_{c} ({\bf r}, t), \qquad \qquad {\bf A}^{(L)} ({\bf r}, t)= {\bf A}_{c}({\bf r}, t).  
\label{eq-B7}
\end{equation}
Similarly, the Coulomb-gauge potentials are obtained by setting $v \rightarrow \infty$ in eq.~(\ref{eq-B1})-(\ref{eq-B4}):
\begin{eqnarray}
\Phi^{(C)}({\bf r},t) = \int {\rho({\bf r}',t) \over |{\bf r} - {\bf r}'|} d^3r', \qquad \qquad \qquad \qquad \qquad \quad \ 
\label{eq-B8}
\\
{\bf A}^{(C)}({\bf r},t) = {\bf A}_{c}({\bf r},t) + c \grad \int \left[ \Phi_{c}({\bf r},t) -  \Phi^{(C)}({\bf r},t) \right] dt.
\label{eq-B9}
\end{eqnarray}

When the parameter $v$ in the velocity gauge is {\em imaginary} in the form of $v = \pm i\nu$ with any {\em real} value of $\nu \ne 0$, the gauge is the generalized $\nu$-Kirchhoff gauge \cite{Yang-Nevels-2014}, with the gauge condition and the equation for the scalar potential listed below:
\begin{equation}
\grad \cdot {\bf A}^{(\nu{\rm K})} - {c \over \nu^2} {\partial \Phi^{(\nu{\rm K})} \over \partial t} = 0,
\label{eq-B10} 
\end{equation}
\begin{equation}
\grad^2 \Phi^{(\nu{\rm K})} + {1 \over \nu^2} {\partial^2 \Phi^{(\nu{\rm K})} \over \partial t^2} = - 4 \pi \rho.
\label{eq-B11}
\end{equation}
When $\nu = c$, the ${\nu}$-Kirchhoff gauge reduces to the original Kirchhoff gauge investigated extensively by Heras \cite{Heras-2006}.  It is obvious that eqs.~(\ref{eq-B10})-(\ref{eq-B11}) are a generalization of Heras's idea of extending $v$ in the velocity gauge to include imaginary values.
Hence, the $\nu$-Kirchhoff scalar potential formally can be expressed as the velocity-gauge scalar potential with an imaginary propagation speed $v = \pm i \nu$. (Ref.\cite{GTM-v-gauge-2022} suggested extending $v$ to the whole complex plane excluding the origin.)

We use a simple example to see what the potential $\Phi^{(\nu {\rm K})} ({\bf r}, t)$ looks like.  We assume a single-frequency charge density of the form:
\begin{equation}
\rho({\bf r}, t) = \rho_{+}({\bf r})e^{i\omega t} + \rho_{-}({\bf r})e^{-i\omega t}, \qquad \qquad \rho_{-}({\bf r}) = [\rho_{+}({\bf r})]^{*},
\label{eq-B12}
\end{equation}
where $^{*}$ denotes complex conjugate.  We assume that the potential has the same time-dependence:
\begin{equation}
\Phi^{(\nu {\rm K})} ({\bf r}, t) = \Phi^{(\nu {\rm K})}_{+}({\bf r})e^{i\omega t} + \Phi^{(\nu {\rm K})}_{-}({\bf r})e^{-i\omega t}, \qquad \qquad 
\Phi^{(\nu {\rm K})}_{-}({\bf r}) = [\Phi^{(\nu {\rm K})}_{+}({\bf r})]^{*}.
\label{eq-B13}
\end{equation}

If we use eqs.~(\ref{eq-B12})-(\ref{eq-B13}) in (\ref{eq-B11}), we have
\begin{equation}
\left( \grad^2 - {\omega^2 \over \nu^2} \right) \Phi_{\pm}^{(\nu {\rm K})} ({\bf r}) = - 4 \pi \rho_{\pm} ({\bf r}).
\label{eq-B14}
\end{equation}
The solution for $\Phi^{(\nu {\rm K})}$ that goes to zero at $|{\bf r}| \to \infty$ is:
\begin{equation}
\Phi^{(\nu {\rm K})} ({\bf r}, t) 
= \int {e^{-|\omega/\nu|R} \over R} \left[  \rho_{+}({\bf r}')e^{i\omega t} + \rho_{-}({\bf r}')e^{-i\omega t} \right] d^3r',
\label{eq-B15}
\end{equation}
where $R = |{\bf r} - {\bf r}'|$.  When both $\omega, \nu \ge 0$, the exponent of the positive-frequency component is: $i\omega t - |\omega/\nu|R = i\omega [t - R/(i\nu)]$, exhibiting an imaginary propagation speed of $i\nu$, first suggested by Heras \cite{Heras-2006}.   (The negative-frequency component is just the complex conjugate of the positive-frequency component.)  We then use this scalar potential in eq.~(\ref{eq-A10}) to get the full expression for the corresponding vector potential ${\bf A}^{(\nu {\rm K})}$.

\section{\bf Gauge transformations of potentials, gauge invariance of fields, and Jackson's gauge-transformation method}
In this section, we show that the solution of the vector potential in eq.~(\ref{eq-A10}) for any gauge has two important properties: (i) any two sets of potentials are related by a gauge transformation, and (ii) the electric and the magnetic fields generated by the potentials are gauge invariant.  

Let us consider two set of potentials ($\Phi$, ${\bf A}$) in eq.~(\ref{eq-A10}) and ($\Phi'$, ${\bf A}'$) listed below:
\begin{equation}
{\bf A}'  ({\bf r}, t) = {\bf A}_{c}  ({\bf r}, t) + c \grad \int \left[ \Phi_{c} ({\bf r}, t) - \Phi' ({\bf r}, t) \right] dt.
\label{eq-G1}
\end{equation}
Next, we define a gauge function $\chi({\bf r}, t)$ by
\begin{equation}
\chi({\bf r}, t) = c \int \left[ \Phi ({\bf r}, t) - \Phi' ({\bf r}, t) \right] dt.
\label{eq-G2}
\end{equation}
Then, these two sets of potentials are related by a gauge transformation via the gauge function $\chi$:
\begin{equation}
{\bf A}' = {\bf A} + c \grad \int \left[ \Phi ({\bf r}, t) - \Phi' ({\bf r}, t) \right] dt = {\bf A} +  \grad \chi,
\label{eq-G3}
\end{equation}
\begin{equation}
\Phi' = \Phi - {1 \over c} {\partial \over \partial t} \left( c \int \left[ \Phi ({\bf r}, t) - \Phi' ({\bf r}, t) \right] dt \right)
= \Phi - {1 \over c} {\partial \chi \over \partial t}.
\label{eq-G4}
\end{equation}
Because of the above two relationships, the electric and the magnetic fields are gauge invariant:
\begin{equation}
{\bf E} = - \grad \Phi' - {1 \over c}{\partial {\bf A}' \over \partial t}
= - \grad \left(\Phi - {1 \over c}{\partial \chi \over \partial t} \right)- {1 \over c}{\partial ( {\bf A} + \grad \chi) \over \partial t} 
= - \grad \Phi - {1 \over c}{\partial {\bf A} \over \partial t},
\label{eq-G5}
\end{equation}
\begin{equation}
{\bf B} = \grad \times {\bf A}' = \grad \times ({\bf A} + \grad \chi) = \grad \times {\bf A}.
\label{eq-G6}
\end{equation}

It is important to note that the potentials $(\Phi, {\bf A})$ in eq.~(\ref{eq-A10}) are just one gauge transformation away from the Lorenz-gauge potentials $(\Phi^{(L)}, {\bf A}^{(L)}) = (\Phi_c, {\bf A}_c)$ via the gauge function $\Lambda_{\rm Jackson}$ defined below:
\begin{equation}
{\bf A}_{\rm Jackson} = {\bf A}^{(L)} + \grad \Lambda_{\rm Jackson}, \qquad \quad  \Lambda_{\rm Jackson} ({\bf r}, t) = c \int \left[ \Phi^{(L)} ({\bf r}, t) - \Phi ({\bf r}, t) \right] dt.
\label{eq-G7}
\end{equation}
This gauge-transformation procedure was first used by Jackson to derive the vector potentials in many gauges of interest \cite{Jackson-2002}.  Thus, the results of Jackson's method totally agree with our universal solution for the vector potential in eq.~(\ref{eq-A10}).  He used the procedure to derive the $v$-gauge vector potential, yielding the same result as in eq.~(\ref{eq-B4}):
\begin{equation}
{\bf A}^{(v)}_{\rm Jackson} = {\bf A}^{(L)} + \grad \Lambda^{(v)}_{\rm Jackson}, \qquad \quad \Lambda^{(v)}_{\rm Jackson} ({\bf r}, t) = c \int \left[ \Phi^{(L)}({\bf r}, t) - \Phi^{(v)}({\bf r}, t) \right]dt.
\label{eq-G8}
\end{equation}
See also Refs.~\cite{Yang-McDonald-2015, Hnizdo-Vaman-2023} for using this procedure in their investigations. 

A close inspection of eq.~(\ref{eq-G7}) reveals that the term $\grad \Lambda_{\rm Jackson}$, which enabled Jackson to go from the Lorenz gauge to an arbitrary gauge, is exactly the vector potential ${\bf A}_2$ listed in eq.~(\ref{eq-A8}).  Hence, eq.~(\ref{eq-A5}) contains all the terms in the Maxwell's equations for potentials responsible for generating the term $\grad \Lambda_{\rm Jackson}$ in Jackson's procedure.

\section{\bf Electromagnetic fields and their propagation in space, and potentials in the Poincaré gauge}

 In this section, we examine the propagation of the fields generated by our potentials. The electric and the magnetic fields ${\bf E}$ and ${\bf B}$ generated by the potentials in an arbitrary gauge, $\Phi$ and ${\bf A}$ in eq.~(\ref{eq-A10}),  are:
 \begin{equation}
{\bf E}  =  - \grad \Phi - {1 \over c}{\partial {\bf A} \over \partial t}
            = - \grad \Phi - \left[ {1 \over c}{\partial {\bf A}_{c} \over \partial t} + \grad (\Phi_{c} - \Phi) \right]
           =  - \grad \Phi_{c} - {1 \over c}{\partial {\bf A}_{c} \over \partial t},
\label{eq-D1}
\end{equation}
\begin{equation}
{\bf B} = \grad \times  \left[ {\bf A}_{c} + c \grad  \int ( \Phi_{c} - 
                       \Phi ) dt \right]
           =  \grad \times {\bf A}_{c}.
\label{eq-D2}
\end{equation}
Because the $c$-retarded potentials $\Phi_c$ and ${\bf A}_c$ always propagate with speed $c$ from the physical charge and current densities $\rho$ and ${\bf J}$ according to eqs.~(\ref{eq-A9}) and (\ref{eq-A6}), eqs.~(\ref{eq-D1})-(\ref{eq-D2}) indicate that the fields always propagate with speed $c$ from the physical charge and current densities $\rho$ and ${\bf J}$.  We note that this statement is true for any scalar potential $\Phi$.

From eqs.~(\ref{eq-D1})-(\ref{eq-D2}), two sets of potentials stand out:~the potentials $(\Phi^{(L)}, {\bf A}^{(L)}) = (\Phi_{c}, {\bf A}_{c})$ in the Lorenz gauge and the potentials ($\Phi^{(G)}$, ${\bf A}^{(G)}$) in the Gibbs gauge \cite{Gibbs-1896, McDonald-2019} (also known as the Hamilton or temporal gauge \cite{Jackson-Okun-2001}):
\begin{equation}
\Phi^{(G)} = 0, \qquad
{\bf A}^{(G)}({\bf r}, t) = {\bf A}_{c}({\bf r}, t) + c \grad  \int \Phi_{c}({\bf r}, t) dt = -c \int {\bf E}({\bf r}, t) dt,
\label{eq-D3}
\end{equation}
\begin{equation}
 - {1 \over c} {\partial {\bf A}^{(G)} \over \partial t} = {\bf E}, 
\qquad
\grad \times {\bf A}^{(G)} = - c \int \grad \times {\bf E} ({\bf r}, t) dt
= \int {\partial {\bf B} ({\bf r}, t) \over \partial t} dt = {\bf B}.
\label{eq-D4}
\end{equation}
We note that ${\bf A}^{(G)}$ is just the time-integration of the local gauge-invariant electric {\em field}.
The vector potential in eq.~(\ref{eq-A10}) also can be expressed as:
\begin{equation}
{\bf A} ({\bf r}, t) 
= {\bf A}^{(G)} ({\bf r}, t) - c \int \grad \Phi ({\bf r}, t) dt
= -c \int \left[ {\bf E} ({\bf r}, t) + \grad \Phi ({\bf r}, t) \right]dt.
\label{eq-D5}
\end{equation}
In particular, the Lorenz-gauge vector potential ${\bf A}^{(L)}$ in eq.~(\ref{eq-B7}) takes our preferred form (sum of a gauge-invariant term and a gauge-dependent term):
\begin{equation}
{\bf A}^{(L)} ({\bf r}, t) = -c \int \left[ {\bf E} ({\bf r}, t) + \grad \Phi^{(L)} ({\bf r}, t) \right]dt.
\label{eq-D6}
\end{equation}
The expression in eq.~(\ref{eq-D5}) for the vector potential is most suitable for discussions of the potentials in the Poincaré gauge next.

The Poincaré or multipolar gauge ({\it e.g.} Refs.~\cite{Jackson-2002, Kobe-MP-1982}) is defined by the gauge condition, ${\bf r} \cdot {\bf A}^{(P)} ({\bf r}, t)= 0$, which is cast in the following form for the discussions of this gauge:
\begin{equation}
\int_0^1 {\bf r} \cdot {\bf A}^{(P)}(u{\bf r}, t) du = 0.
\label{eq-D7}
\end{equation}
The mathematics (but not the physical quantities) used here from eqs.~(\ref{eq-D7}) to (\ref{eq-D12}) closely follows the mathematics in eqs.~(9.1)-(9.9) of Ref.~\cite{Jackson-2002}. If we use eq.~(\ref{eq-D5}), then eq.~(\ref{eq-D7}) takes the form for all $t$,
\begin{equation}
\int_0^1 {\bf r} \cdot \left[ {\bf E}(u{\bf r}, t) +  \grad_{u{\bf r}} \Phi^{(P)}(u{\bf r}, t) \right] du = 0.
\label{eq-D8}
\end{equation}
The integration over $u$ can be done for the scalar potential as follows.  We use the spherical coordinates ${\bf r} = (r, \theta, \phi)$ to get
\begin{equation}
\int_0^1 {\bf r} \cdot \grad_{u{\bf r}} \Phi^{(P)}(u{\bf r}, t) du 
= \int_0^1 r {\partial \Phi^{(P)}(ur,\theta,\phi, t) \over \partial (ur)}  du 
= \Phi^{(P)}({\bf r}, t) - \Phi^{(P)}({\bf 0}, t).
\label{eq-D9}
\end{equation}
Thus, the scalar potential is,
\begin{equation}
 \Phi^{(P)}({\bf r}, t) = -  \int_0^1 {\bf r} \cdot {\bf E}(u{\bf r}, t) du + \Phi^{(P)}({\bf 0}, t).
 \label{eq-D10}
\end{equation}
We then take the gradient of the scalar potential to get (note: $\grad = \grad_{\bf r}$):
\begin{equation}
- \grad  \Phi^{(P)}({\bf r}, t) = {\bf E}({\bf r}, t) - {1 \over c} {\partial \over \partial t} \int_0^1 {\bf r}  \times u{\bf B}(u{\bf r}, t) du.
 \label{eq-D11}
\end{equation}
We now use eq.~(\ref{eq-D11}) in (\ref{eq-D5}) to derive the vector potential,
\begin{equation}
{\bf A}^{(P)}({\bf r}, t) 
= - \int dt \left( {\partial \over \partial t} \int_0^1 {\bf r} \times u{\bf B}(u{\bf r}, t) du \right) 
= - \int_0^1 {\bf r} \times u{\bf B}(u{\bf r}, t) du.
 \label{eq-D12}
\end{equation}

The above discussion shows that our universal vector potential in eq.~(\ref{eq-D5}) or (\ref{eq-A10}) also works for the Poincaré or multipolar gauge.

\section{Conclusions}
In conclusion, we have derived an analytic solution for the vector potential universally valid for any gauge for an arbitrary time-dependent charge-current distribution.  This is done by solving the vector potential directly from Maxwell's equations for potentials without using a gauge condition.  Traditionally, a gauge condition is used to solve for a particular scalar potential.  But as soon as the scalar potential is solved, the corresponding vector potential is completely determined by using the scalar potential in (\ref{eq-A10}) or (\ref{eq-D5}). 

\section{Acknowledgements}

We are most grateful to Professor K.~T.~McDonald for encouragements and for many insightful discussions and suggestions that have greatly enhanced the depth of the paper.  We also thank Professor D.~R.~Wilton for helpful discussions and suggestions.

\appendix 
\section{Appendix: Direct solution of the vector potential in any gauge by Fourier transforms}
\setcounter{equation}{0}
\renewcommand{\theequation}{A.\arabic{equation}}

In this Appendix, we show that the method of Fourier transforms offers a simpler way of obtaining the solution of the vector potential.
Define the Fourier component $\tilde{{\bf A}} ({\bf k}, \omega)$ of the vector potential ${\bf A}({\bf r}, t)$ by:

\begin{equation}
\tilde{{\bf A}} ({\bf k}, \omega) = \int e^{-i {\bf k} \cdot {\bf r} + i \omega t} {\bf A} ({\bf r}, t) d^3r dt,
 \label{eq-APA1}
\end{equation}
\begin{equation}
{\bf A} ({\bf r}, t)  = {1 \over (2 \pi)^4} \int e^{i {\bf k} \cdot {\bf r} - i \omega t} \tilde{{\bf A}} ({\bf k} , \omega) d^3k d\omega,
 \label{eq-APA2}
\end{equation}
and similarly for the scalar potential $\Phi$, {\it etc}.  Thus, equations eqs.~(\ref{eq-A2})-(\ref{eq-A3}) become
\begin{equation}
{\bf k}^2 \tilde{\Phi} 
- {\omega \over c} {\bf k} \cdot \tilde{ {\bf A} } 
= 4 \pi \tilde{ \rho },
\label{eq-APA3}
\end{equation}
\begin{equation}
\left( {\bf k}^2 - {\omega^2 \over c^2} \right) \tilde{ {\bf A} } 
= {4 \pi \over c} \tilde{ {\bf J} } + {\bf k} \left( {\bf k} \cdot \tilde{ {\bf A}} - {\omega \over c} \tilde{\Phi} \right).
\label{eq-APA4}
\end{equation}

Now, we first solve for ${\bf k} \cdot \tilde{{\bf A}}$ from eq.~(\ref{eq-APA3}):
\begin{equation}
{\bf k} \cdot \tilde{{\bf A}} = {c \over \omega} \left( {\bf k}^2 \tilde{\Phi} - 4 \pi \tilde{\rho} \right).
\label{eq-APA5}
\end{equation}
We then use eq.~(\ref{eq-APA5}) in eq.~(\ref{eq-APA4}) to have 
\begin{equation}
\left( {\bf k}^2 - {\omega^2 \over c^2} \right) \tilde{ {\bf A} }  
= {4 \pi \over c} \tilde{ {\bf J} } + {\bf k} \left[ {c \over \omega} \left( {\bf k}^2 \tilde{\Phi} - 4 \pi \tilde{\rho} \right) - {\omega \over c} \tilde{\Phi} \right]
= {4 \pi \over c} \tilde{ {\bf J} } - {4 \pi c \over \omega} {\bf k} \tilde{\rho}
+ {\bf k} {c \over \omega} \left( {\bf k}^2 - {\omega^2 \over c^2} \right) \tilde{\Phi}.
\label{eq-APA6}
\end{equation}
The solution for $\tilde{\bf A}$ is:
\begin{equation}
\tilde{ {\bf A} }  = {4 \pi \over {\bf k}^2 - {\omega^2 / c^2}} \left( {{\bf J} \over c} - c{{\bf k} \over \omega}\tilde{\rho} \right) + c{{\bf k} \over \omega}\tilde{\Phi}
= \tilde{{\bf A}}_{c} + c {i{\bf k} \over - i \omega} (\tilde{\Phi}_{c} - \tilde{\Phi}),
\label{eq-APA7}
\end{equation}
which is exactly eq.~(\ref{eq-A10}).

\section{Appendix: Direct solution of the vector potential from Maxwell's equations for potentials in the velocity gauge}
\setcounter{equation}{0}
\renewcommand{\theequation}{B.\arabic{equation}}

In this Appendix, we present one method for solving the vector potential {\em directly} from Maxwell's equations for potentials {\em in the velocity gauge}.  In the velocity gauge, the gauge condition is in eq.~(\ref{eq-B1}), the equation for the scalar potential is in eq.~(\ref{eq-B2}) and the equation for the vector potential is:
\begin{eqnarray}
\left( \grad^2 - {1 \over c^2} {\partial^2 \over \partial t^2}  \right) {\bf A} ^{(v)} 
= - { 4 \pi \over c } {\bf J} + c \grad \left( - {1 \over v^2} {\partial \over \partial t} + {1 \over c^2} {\partial  \over \partial t} \right) \Phi^{(v)} \ \ 
\nonumber
\\
= - { 4 \pi \over c } {\bf J} + c \grad \int \left( - {1 \over v^2} {\partial^2  \over \partial t^2} + {1 \over c^2} {\partial^2  \over \partial t^2} \right) \Phi^{(v)}  dt
\qquad \qquad \qquad \quad \ \ \,
\nonumber
\\
= - { 4 \pi \over c } {\bf J} + c \grad \int \left[ \left(\grad^2 - {1 \over v^2} {\partial^2 \over \partial t^2}\right) 
                           - \left(\grad^2 - {1 \over c^2} {\partial^2  \over \partial t^2} \right) \right] \Phi^{(v)} dt
\ \ \,
\nonumber
\\
= - { 4 \pi \over c } {\bf J} - 4 \pi \left( c \grad \int \rho dt \right)
                         - \left(\grad^2 - {1 \over c^2} {\partial^2  \over \partial t^2} \right) \left(c \grad \int \Phi^{(v)} dt \right).
\label{eq-APB1}
\end{eqnarray}
It is obvious that the solution for eq.~(\ref{eq-APB1}) is exactly the vector potential listed in eq.~(\ref{eq-B4}).

\end{document}